# The reason for the tilting of domain wall with Dzyaloshinskii-Moriya interaction from a microscopic dynamical perspective

Maokang Shen[1], Yue Zhang[1*], Wei Luo[1], Long You[1], Xiaofei Yang[1]

[1]*School of Optical and Electronic Information, Huazhong University of Science and Technology, Wuhan, 430074, PR China*

* Corresponding author. **Electronic mail:** yue-zhang@hust.edu.cn

**Abstract**

The interfacial Dzyaloshinskii-Moriya interaction (DMI) of a heavy metal (HM)/ferromagnetic (FM) metal heterostructure is vital to the current-induced domain wall motion (CIDWM) at an ultrahigh velocity. However, strong DMI also tilts the moving domain wall (DW) plane, and the mechanism for this tilting is not quite clear. In this work, we have found that this tilting may be understood based on a micromagnetic calculation from a microscopic dynamical perspective. The DMI-induced *antisymmetric* moment structure at the two boundaries of the track needs to be paid attention. In the early stage of CIDWM induced by spin-orbit torque, this antisymmetry is destroyed. Afterwards, the moments at the two boundaries experience distinct rotation processes with different energy paths towards their final stable antisymmetric moment structure. This results in different initial velocities of the local DW regions at the two boundaries. In mathematics, this distinct DW dynamical progresses at the two boundaries can be approximately revealed by modifying the initial conditions for solving the Thiele equations.

The racetrack memory based on the current-induced-domain-wall-motion (CIDWM) has potential for developing novel magnetic memory device with a high reading speed [1]. When compared with the CIDWM by spin-transfer-torque effect in a ferromagnetic metal (FM) nanowire, the CIDWM by spin-orbit-torque (SOT) effect in a heavy metal (HM)/FM bilayer nanowire exhibits a higher energy efficiency [2-4]. The DW in the HM/FM bilayer can be driven at a very low current density. This striking advantage is relevant to an antisymmetric exchange coupling, the Dzyaloshinskii-Moriya interaction (DMI), at the HM/FM interface [4]. The DMI results in a Néel-typed chiral DW and offers a strong torque that switches the DW moments and induces the DW to move at a high speed [5]. On the other hand, strong DMI also reshape the moving DW. The DW plane with strong DMI tilts when the DW are driven under SOT or a magnetic field [6-8]. The tilting of DW plane limits the increase in the storage density in a real racetrack memory [9].

Initially, the DMI-related DW tilting is ascribed to the competition between DMI and SOT. The SOT tends to rotate DW moments, but the DMI pins it. As a compromise, the whole DW plane tilts at a price of the increased shape anisotropy energy [7]. Based on this theory, Boulle et al depict the DW tilting using the CCM with the tilting angle as an added collective coordinate [7]. In recent years, more attention is paid on the difference in the edge energies at the two boundaries of the track. Considering the edge energy, Muratov et al prove an energically stable straight tilting DW plane under an external magnetic field when DMI is strong enough [10]. However, Kim et al believe that for a curved DW plane, the DW tilting is governed by the speed asymmetry due to the difference of the total effective fields at the two boundaries but not simply by DMI [11]. Additionally, the shape of the medium seems also matters to the DW tilting. The chirality-independent DW tilting is observed in the curved track [12]. In a circular dot, Baumgartner et al observe a long-term (> 1 ns) evolution of the DW tilting after the initial nucleation [13]. Because of this long-term evolution, one may see different directions of DW tilting by using different modes (static or time-resolved) for the observation [14]. Therefore, the mechanism for the DMI-related DW tilting is still not quite clear. Especially, the generation of the tilting from a microscopic dynamical perspective is not ascertained.

In this work, using micromagnetic simulation and calculation, the evolution of the DW tilting is studied from a microscopic perspective. Without considering the extrinsic factors such as medium morphology and defect, the tilting of DW plane in a long track is intrinsically related to the *antisymmetric* moment structure at the two track boundaries. This antisymmetry leads to distinct switching of the DW moments at the two boundaries, resulting in different initial velocities of the local DW at the two boundaries. This evolution of DW tilting can be accomplished in a very short time (much shorter than 1 ns). Therefore, the DW tilting may not depend on the observation modes.

The simulation was done by the software named the Object-Oriented MicroMagnetic Framework (OOMMF) containing the codes of the damping-like SOT and DMI [15]. We considered the CIDWM in a 100-nm-wide, 2000-nm-long, and 0.6-nm-thick track. The size of unit cell was 1 nm (length) × 1 nm (width) × 0.6 nm (thickness). The parameters of Co/Pt, a typical ultra-thin film with perpendicular magnetic anisotropy (PMA) and strong DMI, were used for the simulation [9, 16]: the saturation magnetization $M_S = 7 \times 10^5$ A/m, the PMA constant $K = 8 \times 10^5$ J/m$^3$, the exchange stiffness constant $A = 1 \times 10^{11}$ J/m, the DMI constant $D = -1.5$ mJ/m$^2$, the current density $J = 5 \times 10^{11}$ A/m$^2$, the spin Hall angle $\theta_{SH} = 0.08$, and the damping coefficient $\alpha = 0.03$. The $\alpha$ is smaller than the damping coefficient for Co/Pt film [17]. We chose this small $\alpha$ for prolonging the short-term initial relaxation in order to reveal the early evolution of the DW tilting in detail.

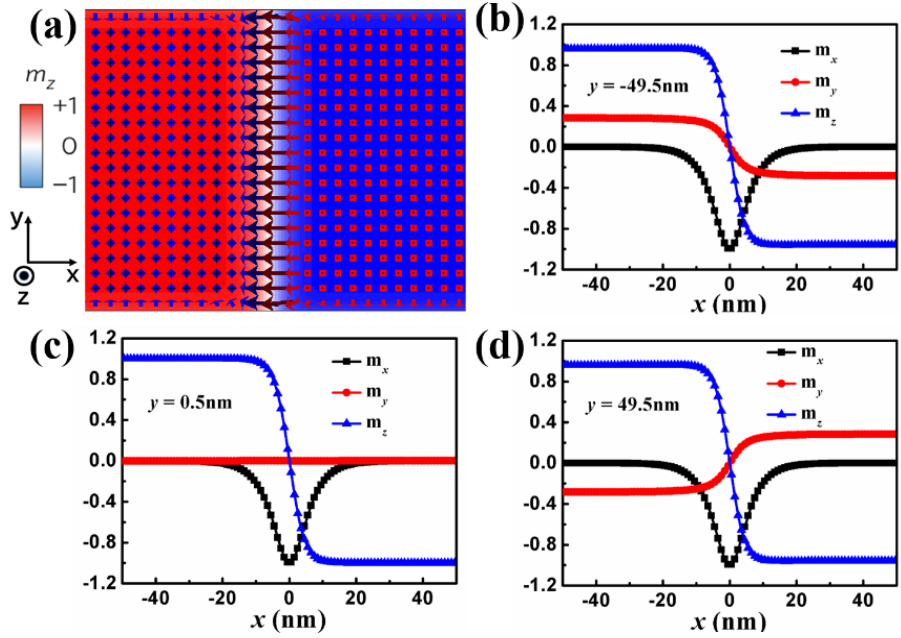

**Figure 1. The moment structure of the left-handed Néel-typed DW (a), the variation of $m_x$, $m_y$, and $m_z$ across the DW region at the lower boundary of the track (b), in the inner part of the track (c), and at the upper boundary of the track (d).**

Initially, a stable DW was generated in the middle of the track. Because of the negative $D$, the DW exhibits a Néel-typed structure with left-handed chirality (Fig. 1(a)). In the inner part of the track (Fig. 1(c)), the moments rotate in the $xz$ plane across the DW region. On the other hand, near the two boundaries, a small projection of the moment on the $y$ axis was found due to the DMI-related boundary condition [15]. It is noteworthy that the direction for the moment projection at one boundary is opposite to that at the other one, resulting in the antisymmetric moment structures (Fig. 1(b) and (d)). However, this moment projection on the $y$ axis makes very small impact on the $x$ dependence of $m_x$ and $m_z$.

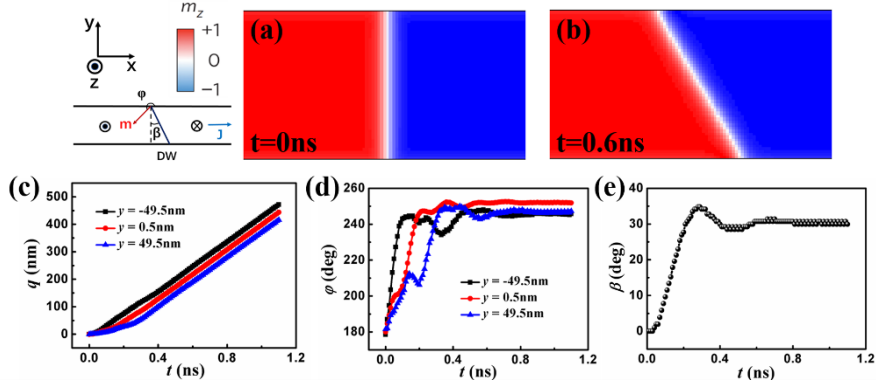

**Figure 2. The snapshots of the DW at the initial state (a) and after injecting the current for 0.6 ns (b). The time-dependence of the central position (c), the azimuthal angle (d), and the tilting angle (e) of DW at the two boundaries and in the inner part of track.**

The CIDWM was found after the current is injected due to the combined action of SOT and DMI. After moving for 0.6 ns, a straight DW with the tilting angle of around 30 degree was generated (Fig. 2(b)). The dynamics of the motion and tilting of DW may be described by several collective coordinates, including the central position $q$, the azimuthal angle $\varphi$ of the moment in the central of the DW, and the tilting angle $\beta$ for the DW plane [7, 8]. In previous papers, these collective coordinates were used to depict the whole DW without considering the microscopic moment structure in local DW region. In this work, we found different evolution of $q$ and $\varphi$ between in the inner part and near the boundaries in the early stage of CIDWM. The DW region at the upper boundary moves clearly faster than that at the lower one, and they share the same velocity finally (Fig. 2(c)). On the other hand, the moment in the central of DW at the lower boundary rotates much faster than that at the upper one, and they rotate to the same direction eventually (Fig. 2(d)). The tilting angle of the DW plane becomes constant after the DW velocities and the boundary azimuthal angles become stable (Fig. 2(e)). This indicates that the *intrinsic* DW tilting is attributed to the difference of the DW motion between at the upper and the lower boundaries.

What needs to be paid attention is that the initial moment states in the middle of the DW at the boundary and in the inner part of the track are totally the same (Fig. 1 (b)-(d)). This offers the *same initial conditions* for solving the Thiele equations derived in the previous paper [7, 8]. Therefore, the collective coordinate method (CCM) exploited in the previous work may be inadequate to reveal the CIDWM at the track boundary. To understand the DW tilting, the DMI-induced antisymmetric moment structure at the two boundaries must be considered in the initial conditions.

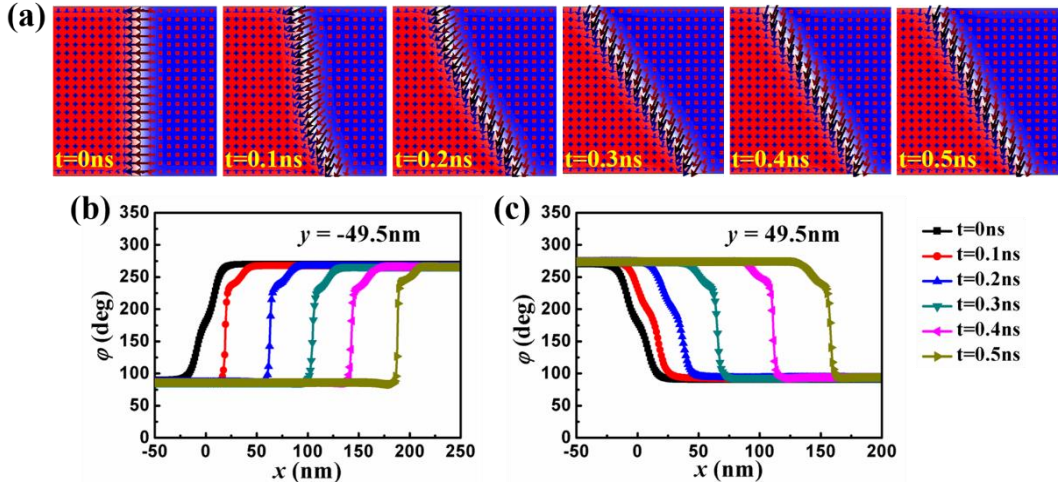

**Figure 3. The snapshots depicting the early evolution of the DW tilting (a), and the spatial distribution of the azimuthal angle across the DW at the lower (b) and upper (c) boundary of the track in the initial 0.5 ns after injecting the current.**

To ascertain the detailed process of DW motion at the two boundaries, the evolution of the local moment structure near the boundaries was collected from the simulation result (Fig. 3). One can see that the antisymmetric moment structure at the two boundaries is destroyed in the early stage of the DW motion. When $t = 0.1$ ns, the moments in the central of the DW at the lower boundary are greatly rotated, resulting in a sharp transition of moment orientation across the DW. However, at the upper boundary, the variation of moment orientation across the DW is still very smooth at $t = 0.1$ ns and 0.2 ns. When $t = 0.5$ ns, the moment structure at the two boundaries becomes antisymmetric again, which is accompanied with the identical DW velocities at the two boundaries.

The DW velocity at the boundary is mainly attributed to the edge energy. Based on the moment structure shown in Fig. 3, the edge free energy density including anisotropy energy, demagnetization energy, exchange energy, and the DMI energy at 0.1 and 0.5 ns were all derived (Fig. 4). The DMI and demagnetization energy density at the two boundaries are quite different at 0.1 ns. At the upper boundary, the lower DMI and demagnetization energy offer a trap that pins the reorientation of moments and the CIDWM. When $t = 0.5$ ns, however, the energy densities at the two boundaries become the same, resulting in the same DW velocity.

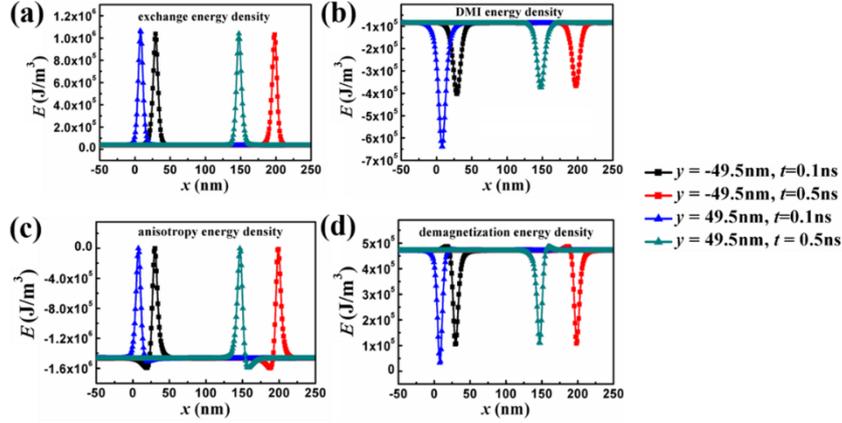

**Figure 4.** The energy densities for the exchange energy (a), the DMI energy (b), the anisotropy energy (c), and the demagnetization energy (d) at the upper and lower boundaries when $t =$ 0.1 ns and 0.5 ns.

The DW motion is factually the transfer of moment switching along the track length direction. In the present system, the current flowing along the $x$ direction induces a SOT that rotates the moments towards $-y$ direction. Afterwards, the moments are switched to the $z$ direction by the torque from DMI effective field [5]. This process is equal to moving the DW to the right. This is the overall process for the DW motion. However, the moment switching experiences different paths at the two boundaries. In the early stage of DW motion, what determines the initial velocity of the DW motion is not the orientation of moment in the central of DW but that at the right side of DW since the DW moves to the right. At the two boundaries, the moments at the right side of the DW plane tilt to opposite directions in their initial states (Fig. 1). When the current is injected, the moments at the two boundaries experience distinct paths to reach their final state (Fig. 3). The moments at the right side of DW plane at the lower boundary are easy to be rotated to the $-y$ direction by SOT and their final stable state can be reached in a short time (Fig. 3(a) and (b)), which results in a large local velocity. On the contrary, at the upper boundary, the moments at the right side of DW need to be rotated by more degrees to escape the energy trap of DMI and demagnetization when the moment is along the $-x$ direction. As a result, the initial DW motion at the upper boundary at 0.1 ns is slower than that at the lower one. This different initial velocity for the upper and lower DW moments results in different time to reach their final antisymmetric states.

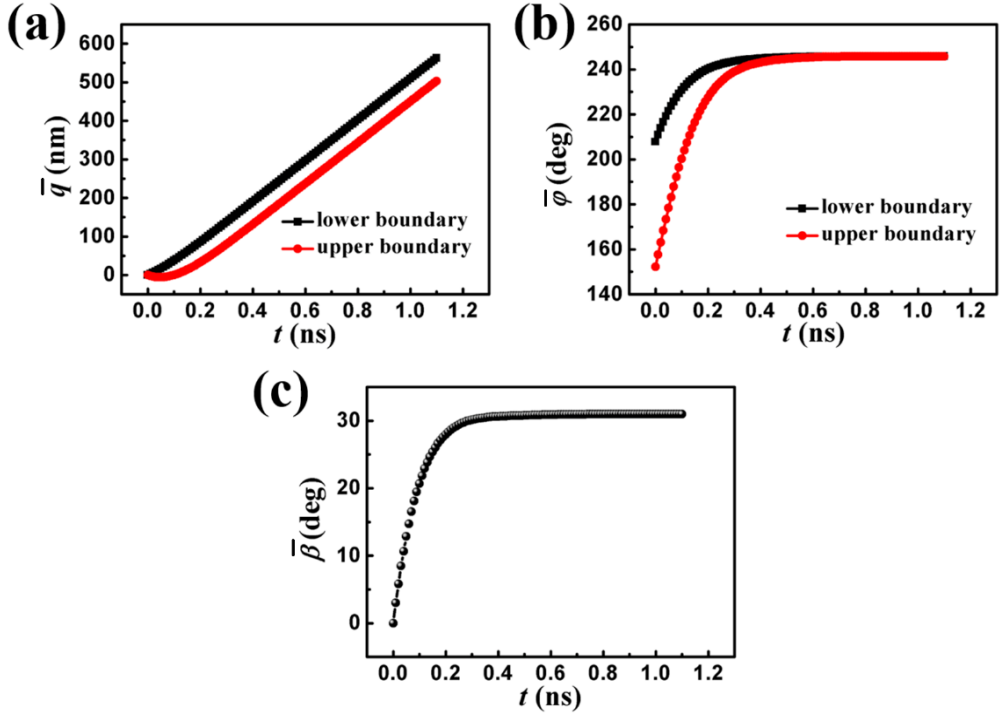

**Figure 5. The temporal central position (a), the azimuthal angle (b), and the tilting angle (c) by averaging the solutions of the CCM equations by taking the orientation of the moments at the right side of the DW as the initial condition.**

From a mathematic point, the above distinct physical progress at the two boundaries can be approximately taken as solving a dynamical equation with different initial conditions. For a Néel-typed chiral DW, the initial $\varphi$ is generally taken as 0 or 180 degree, which is true for the moments across the DW in the inner part of track (Fig. 1(c)). Nevertheless, the initial $\varphi$ at the boundaries varies with $x$ across the DW (Fig. 1(b) and (d)), and the initial $\varphi$ needs to be modified. In this work, a series of $\varphi$ for the moment at the right side of DW was used for the initial condition. It varies from 180 degree to 90 and −90 degree for the moments at the upper and lower boundaries, respectively. For every initial $\varphi$ at each boundary, the temporal $q$ and $\varphi$ were determined numerically by solving the Thiele equations

$$\frac{\alpha}{\Delta}\dot{q} + \dot{\varphi} = \frac{\mu_0 \gamma \pi H_{SO} J \cos\varphi}{2} \tag{1},$$

$$\frac{1}{\Delta}\dot{q} - \alpha\dot{\varphi} = -\frac{\gamma \pi D \sin\varphi}{2\Delta M_S} - \mu_0 \gamma N_x M_S \sin\varphi \cos\varphi \tag{2}.$$

Here, $\alpha$, $\gamma$, $\mu_0$, $N_x$, and $J$ are the damping coefficient, the gyromagnetic ratio of an electron, the permeability of vacuum, the demagnetization factor, and the current density. The $H_{SO}$ is the effective magnetic field for the damping-like SOT and is expressed as

$$H_{SO} = \frac{\mu_B \theta_{SH}}{\mu_0 \gamma e M_s L_z} \quad (3).$$

Here $\mu_B$, e, and $L_z$ are the Bohr magneton, the charge of an electron, and the thickness of the FM film, respectively.

Finally, all the $q$ and $\varphi$ values were averaged to represent the DW position and the azimuthal angle for the moments at the two boundaries. The tilting angle was determined from the difference of the stable $q$ at the two boundaries. The results shown in Fig. 5 indicate that after the early DW motion with different velocities at the two boundaries, their final velocities become the same. This is accompanied by the different early change of $\varphi$. The average $q$, $\varphi$, $\beta$, and the time for this initial progress are all consistent well the simulation result.

In a real track, in addition to DMI, the morphology of the medium, and the pinning due to the edge defects may also influence the tilting of DW [11, 13, 14]. However, in the present work, we found that the DMI itself indeed can lead to the DW tilting in a very short time. In a real Pt/Co track, this process may occur even more quickly because of its larger damping coefficient. However, to observe such an *intrinsic* DW tilting is challenging in experiment. In the reported observation of DW tilting in the track with tens of width or in the circular dot, this intrinsic function of DMI may be concealed by the extrinsic factors. We need to prepare a narrow track that is pure enough, and observe the DW tilting through an ultra-fast technique.

In summary, to understand the mechanism for the tilting of a CIDWM with strong DMI, the DMI-related antisymmetric moment structure at the two boundaries needs to be considered. This antisymmetry was destroyed in the early stage of the unidirectional CIDWM. For a left-handed DW moving to the right, the moments at the right side of DW at the lower boundary are easy to rotate, resulting in a higher initial velocity. At the upper one, the moments at the right side of DW needs to escape the energy traps of DMI and demagnetization, leading to a lower initial velocity. This velocity difference disappears finally when the moment structures at the two boundaries become antisymmetric again. In mathematics, this different dynamical progresses can be approximately seen as solving the same dynamical equation with different initial conditions.

**Acknowledgements**

This work was supported by the National Natural Science Foundation of China [grant number 11574096] and Huazhong University of Science and Technology (No. 2017KFYXJJ037).